\documentstyle[aps,prl,multicol,epsf]{revtex}
\begin{document}
\title{Anomalous temperature behavior of resistivity in lightly doped
manganites around a metal-insulator phase transition}
\author{S. Sergeenkov$^{1,2}$, M. Ausloos$^{1}$, H. Bougrine$^{1,3}$, A.
Rulmont$^{4}$, R. Cloots$^{4}$}
\address{$^{1}$SUPRAS, Institute of Physics, B5, University of Li$\grave e$ge,
B-4000 Li$\grave e$ge, Belgium\\ $^{2}$Bogoliubov Laboratory of
Theoretical Physics, Joint Institute for Nuclear Research,\\ 141980
Dubna, Moscow Region, Russia\\ $^{3}$SUPRAS, Montefiore Electricity
Institute, B28, University of Li$\grave e$ge, B-4000 Li$\grave e$ge,
Belgium\\ $^{4}$SUPRAS and LCIS, Institute of Chemistry, B6,
University of Li$\grave e$ge, B-4000 Li$\grave e$ge, Belgium\\} \draft
%\preprint{to appear in JETP Lett.}
\maketitle
\begin{abstract}
An unusual temperature behavior of resistivity $\rho (T,x)$ in
$La_{0.7}Ca_{0.3}Mn_{1-x}Cu_xO_3$ has been observed at slight $Cu$
doping ($0\leq x \leq 0.05$). Namely, introduction of copper results
in a splitting of the resistivity maximum around a metal-insulator
transition temperature $T_0(x)$ into two differently evolving peaks.
Unlike the original $Cu$-free maximum which steadily increases with
doping, the second (satellite) peak remains virtually unchanged for
$x<x_c$, increases for $x\ge x_c$ and finally disappears at $x_m\simeq
2x_c$ with $x_c\simeq 0.03$. The observed phenomenon is thought to
arise from competition between substitution induced strengthening of
potential barriers (which hamper the charge hopping between
neighboring $Mn$ sites) and weakening of carrier's kinetic energy. The
data are well fitted assuming a nonthermal tunneling conductivity
theory with randomly distributed hopping sites.
\end{abstract}

\begin{multicols}{2}

To clarify the underlying microscopic transport mechanisms in
exhibiting colossal magnetoresistance manganites, numerous studies
(both experimental and theoretical) have been undertaken during the
past few years~\cite{1,2,3,4,5,6,7,8,9,10,11,12,13,14,15,16,17} which
revealed a rather intricate correlation of structural, magnetic and
charging properties in these materials based on a crucial role of the
$Mn^{3+}-O-Mn^{4+}$ network. In addition to the so-called
double-exchange (DE) mechanism (allowing conducting electrons to hop
from the singly occupied $e_{2g}$ orbitals of $Mn^{3+}$ ions to empty
$e_{2g}$ orbitals of neighboring $Mn^{4+}$ ions), these studies
emphasized the important role of the Jahn-Teller (JT) mechanism
associated with the distortions of the network's bond angle and length
and leading to polaron formation and electron localization in the
paramagnetic insulating region. In turn, the onset of ferromagnetism
below Curie point increases the effective bandwidth with simultaneous
dissolving of spin polarons into band electrons and rendering material
more metallic. To modify this network, the substitution effects on the
properties of the most popular $La_{0.7}Ca_{0.3}MnO_3$ manganites have
been studied including the isotopic substitution of oxygen ("giant"
isotope effect~\cite{8,9}), rare-earth (RE)~\cite{10,11,12,13,14} and
transition element (TE)~\cite{15,16,17} doping at the $Mn$ site. In
particular, an unusually sharp {\it decrease} of resistivity $\rho
(T)$ in $La_{0.7}Ca_{0.3} Mn_{0.96}Cu_{0.04}O_3$ due to just $4\%$
$Cu$ doping has been reported~\cite{17} and attributed to the $Cu$
induced weakening of the kinetic carrier's energy $E_0(x)$. On the
other hand, the opposite temperature behavior of resistivity (that is
an {\it increase} of $\rho $ upon TE doping) can also be expected
based on deactivation of the DE Zener mechanism. Indeed, this
mechanism is effective when electrons can hop (tunnel) between
nearest-neighbor TE ions without altering their spin or energy. Hence,
the observed~\cite{16} lowering of the metal-insulator (M-I)
transition temperature and hopping based conductivity by TE
substitution can be ascribed to an inequivalence of the ground-state
energies of neighboring $Mn$ and TE ions resulting in an appearance of
the doping dependent potential barrier $U(x)$. More
precisely~\cite{15,16}, this potential energy exceeds the polaron
bandwidth (virtually weakening the DE interaction between neighboring
TE and $Mn$ ions and impeding thus the possibility of
energy-conserving coherent hops) and is defined as the difference
between the binding energies of an electron on a TE ion (e.g., $Cu$)
and $Mn$ ion, respectively.

In an attempt to pinpoint the above-mentioned potential energy
controlled hopping mechanism and gain some insight into the barrier's
doping profile, in this Letter we present a comparative study of
resistivity measurements on $Cu$ doped polycrystalline manganite
samples from the $La_{0.7}Ca_{0.3} Mn_{1-x}Cu_xO_3$ family for $0\leq
x\leq 0.05$ for a wide temperature interval (from $20K$ to $300K$). As
we shall see, the data are reasonably well fitted (for all $T$ and
$x$) by a unique (nonthermal) tunneling expression for the resistivity
assuming a random (Gaussian) distribution of hopping sites and an
explicit form for the temperature and doping dependent effective
potential $U_{eff}(T,x)=U(x)-E(T,x)$. Besides, the $Cu$ doping induced
competition between the barrier's height profile $U(x)$ and the
previously found~\cite{17} behavior of the carrier's kinetic energy
$E_0(x)\equiv E(0,x)$ results in emergence of a satellite peak in the
temperature behavior of the observed resistivity on the insulating
side.

The samples examined in this study were prepared by the standard
solid-state reaction from stoichiometric amounts of $La_2O_3$,
$CaCO_3$, $MnO_2$, and $CuO$ powders. The necessary heat treatment was
performed in air, in alumina crucibles at $1300C$ for 2 days to
preserve the right phase stoichiometry. Powder X-ray diffraction
patterns are characteristic of perovskites and show structures that
reflect the presence of orthorhombic (or tetragonal) distortions
induced by $Cu$ doping. It was confirmed that our data for the undoped
samples are compatible with the best results reported by other groups
ensuring thus the quality of our sample processing conditions and
procedures.

The electrical resistivity $\rho (T,x)$ was measured using the
conventional four-probe method. To avoid Joule and Peltier effects, a
dc current $I=1mA$ was injected (as a one second pulse) successively
on both sides of the sample. The voltage drop $V$ across the sample
was measured with high accuracy by a KT256 nanovoltmeter. Figure 1
presents the temperature behavior of the resistivity $\rho (T,x)$ for
six $La_{0.7}Ca_{0.3}Mn_{1-x}Cu_xO_3$ samples, with $0\leq x\leq
0.05$. Notice a rather broad bell-like form of resistivity for the
undoped sample (Fig.1(a)) reaching a maximum at the so-called
metal-insulator transition (peak) temperature $T_0(0)=200K$. Upon $Cu$
doping, two markedly different processes occur. First of all, the
$Cu$-free (left) resistivity peak increases and becomes more narrow
(with $T_0(x)$ shifting towards lower temperatures). Secondly, at a
higher temperature another (satellite) peak emerges splitting from the
original one. It remains virtually unchanged for small $x$ (up to
$x_c\simeq 0.03$) and starts to increase for $x>x_c$ until it finally
merges with the main (left) peak at the highest doping level of
$x=0.05$.

Due to tangible microstructural changes (observed upon copper doping),
the JT mechanism plays a decisive role in the above-described
resistivity anomalies by assisting electron localization near the M-I
transition temperature. Given the growing experimental
evidence~\cite{14,15} that polaronic distortions (evident in the
paramagnetic state) persist in the ferromagnetic phase as well, we
consider the observed resistivity to arise from tunneling of small
spin polarons through the doping created potential barriers. According
to a conventional picture~\cite{5,6,7,14,17}, the conductivity due to
tunneling of a carrier through an effective barrier of height
$U_{eff}$ and width $R$ reads
\begin{equation} \sigma =\sigma _he^{-2R/L},
\end{equation}
where $L=h/\sqrt{2mU_{eff}}$ is a characteristic length with $h$ the
Plank's constant and $m$ an effective carrier mass.

To account for the observed anomalous behavior of the resistivity in
our samples, we assume that around the metal-insulator transition in
addition to the Cu-doping induced slight modification ($x\ll 1$) of
the barrier's height $U(x)\equiv U_{eff}(T_0,x) \simeq xU_1+(1-x)U_2$,
the effective potential $U_{eff}(T,x)=U(x)-E(T,x)$ will also depend on
the temperature via the corresponding dependence of the carrier's
energy $E(T,x)=h^2/2m\xi ^2(T,x)$ with some characteristic length $\xi
(T,x)\simeq \xi _0(x)/[1-T/T_0(x)]$ (with~\cite{17} $\xi _0(x)\simeq
\xi _0(0)/(1-x)^2$) which plays a role of the charge carrier
localization length above $T_0$ (in insulating phase) and the
correlation length below $T_0$ (in metallic phase), so that $\xi
^{-1}(T_0,x)=0$. Furthermore, given a rather wide temperature
dependence of resistivity for the undoped sample (see Fig.1(a)), we
adopt the effective medium approximation scheme and assume a random
distribution of hopping distances $R$ with the normalized function
$f(R)$ leading to
\begin{equation}
\rho \equiv <\sigma ^{-1}>=\frac{1}{Z}\int_0^{R_m}dRf(R)\sigma
^{-1}(R),
\end{equation}
for the effective medium resistivity, where $Z=\int_0^{R_m}dRf(R)$
with $R_m$ being the largest hopping distance. In what follows, for
simplicity we consider a Gaussian distribution (around a mean value
$R_0$) with the normalized function $f(R)=(2\pi
R_0^2)^{-1/2}e^{-R^2/2R_0^2}$ resulting in the following expression
for the observed resistivity
\begin{equation}
\rho (T,x)=\rho _h e^{\gamma ^2}\left [\frac{\Phi (\gamma )- \Phi
(\gamma -\gamma _m)}{\Phi (\gamma _m)}\right ],
\end{equation}
where
\begin{equation}
\gamma (T,x)=\sqrt{\mu (x)-\gamma _0(x)\left [1-\frac{T}{T_0(x)}\right
]^2},
\end{equation}
with
\begin{equation}
\mu (x)=\frac{2mU(x)R_0^2}{h^2}\equiv \mu (0)+x\Delta \mu ,
\end{equation}
(which measures the substitution induced potential barriers $U(x)$
hampering the charge hopping between neighboring $Mn$ sites) and
\begin{equation}
\gamma _0(x)=\frac{R_0^2}{\xi _0^2(x)}\simeq \gamma _0(0)(1-x)^4,
\end{equation}
(which measures the effects due to the carrier's kinetic energy
$E_0(x)\equiv E(0,x)$, see above). Here $\rho _h=1/\sigma _h$, $\gamma
_m=R_m/R_0$, and $\Phi (\gamma )$ is the error function.

Turning to the discussion of the main (left) resistivity profile, we
note that the $Cu$ induced changes of its peak temperature $T_0(x)$
are well fitted by the exponential law
\begin{equation}
T_0(x)=T_0(0)-T_m\left (1-e^{-x\tau }\right ),
\end{equation}
with $T_0(0)=200K$, $T_m=73K$ and $\tau=56$. At the same time,
according to Eqs.(3)-(7) (and in agreement with the observations, see
Fig.2), the corresponding peak resistivity $\rho _0(x)\equiv \rho
(T_0,x)$ increases with $x$ as follows
\begin{equation}
\rho _0(x)=\rho _0(0)e^{x\Delta \mu },
\end{equation}
yielding $\rho _0(0)=\rho _he^{\mu (0)}=4m\Omega m$, and $\Delta \mu
=54$ for the model parameters and suggesting that $\rho _0(x)\propto
1/T_0(x)$. To further emphasize this similarity, Fig.2 depicts the
extracted doping variation of the normalized quantities,
$[T_0(0)-T_0(x)]/T_m$ (open dots) and left peak conductivity $\sigma
_0(x)/\sigma _0(0)=\rho _0(0)/\rho _0(x)$ (solid dots) along with the
fitting curves (solid lines) according to Eqs.(7) and (8).

A more careful analysis of Eq.(3) shows that in addition to the main
peak at $T_0(x)$, equation $\frac{d\rho (T,x)}{dT}=0$ has two more
conjugated extreme points at $T=T_S^{\pm}(x)$ intrinsically linked to
the main peak, viz. $T_S^{-}(x)=T_0(x)\left [1-\sqrt{\frac{\mu (x)-\mu
^{-}}{\gamma _0(x)}} \>\right ]$ and $T_S^{+}(x)=T_0(x)\left
[1-\sqrt{\frac{\mu ^{+}-\mu (x)}{\gamma _0(x)}} \> \right ]$ with $\mu
^{\pm}=\sqrt{2}(2\pm \gamma _m)$. To attribute these temperatures to
the observed satellite (right) peak (see Fig.1), first of all, we have
to satisfy the "boundary conditions" at zero ($x=0$) and highest
($x=x_m=0.05$) doping levels by assuming $T_S^{-}(0)=T_0(0)$ and
$T_S^{+}(x_m)=T_0(x_m)$ which lead to the following constraints on the
model parameters: $\mu ^{-}=\mu (0)$ and $\mu ^{+}=\mu (x_m)$. And
secondly, to correctly describe the observed evolution of the
satellite peak with copper doping and to introduce a critical
concentration parameter $x_c$ into our model, we use the continuity
condition $T_S^{+}(x_c)=T_S^{-}(x_c)$. As a result, we find that the
satellite's peak is governed by the unique law over the whole doping
interval with
\begin{eqnarray}
T_S^{-}(x)&=&T_0(x)\left [1-\sqrt{\frac{x}{2x_c}} \ \right ], \quad
\qquad 0\le x<x_c,
\\ T_S^{+}(x)&=&T_0(x)\left [1-\sqrt{1-\frac{x}{2x_c}} \ \right
], \quad x_c\le x \le x_m,
\end{eqnarray}
where $x_m=2x_c$ with $x_c=\gamma _0(0)/\Delta \mu$. Noting that
according to Eqs.(4)-(10), $\gamma ^2(T_S^{-})=\mu (0)$ and $\gamma
^2(T_S^{+})=\mu (0)+2(x-x_c)\Delta \mu$, in good agreement with the
observations (see Fig.1) it follows now from Eq.(3) that indeed the
satellite peak shows practically no changes with $x$ (up to $x\simeq
x_c$) since $\rho _S^{-}(x)=\rho _h\exp[\gamma ^2(T_S^{-})]\simeq \rho
_0(0)$ and starts to increase above the threshold (for $x>x_c$) as
$\rho _S^{+}(x)=\rho _0(0)\exp[2(x-x_c)\Delta \mu ]$ until it totally
merges with the main peak at $x\simeq x_m$. By comparing the above
expressions with our experimental data for resistivity peaks at
$x=x_c$ and $x=x_m$, we get $\gamma _0(0)=R_0^2/\xi _0^2(0)\simeq 1.5$
which (along with extracted above value of $\Delta \mu$, see Eq.(8))
leads to $x_c=\gamma _0(0)/\Delta \mu =E_0(0)/\Delta U\simeq 0.03$ for
the critical concentration of copper, in very good agreement with the
observations. As expected, $x_c$ reflects the competition between the
carrier's kinetic energy and the copper induced potential barrier. In
turn, assuming as usual~\cite{5,6,14,17} $R_0\simeq 5.5\AA$ for a mean
value of the hopping distance and using a free-electron mass value for
$m$, the above estimates yield $U_2\equiv U(0)\simeq E_0(0)\simeq 0.1
eV$ and $U_1 \simeq \Delta U\simeq 3 eV$ for the barrier's height of
the undoped and maximally doped samples, respectively.

Finally, given the above explicit dependencies for $T_0(x)$ and
$T_S^{\pm}(x)$ along with the fixed model parameters, we are able to
fit {\it all} the resistivity data with a single function $\rho (T,x)$
given by Eq.(3). The solid lines in Fig.1 are the best fits according
to this equation assuming nearest-neighbor hopping approximation (with
$\gamma _m=1$).

In summary, due to the competition between the copper modified kinetic
carrier's energy $E(0,x)$ and the potential barriers $U(x)$ between
$Mn^{3+}-Mn^{4+}$ dominated hopping sites, a rather unusual
"double-peak" behavior of the resistivity $\rho (T,x)$ is observed in
$La_{0.7}Ca_{0.3}Mn_{1-x}Cu_xO_3$ at slight $Cu$ doping around the
metal-insulator transition temperature $T_0(x)$. The temperature and
$x$ dependencies of the resistivity are rather well fitted by a
coherent (nonthermal) tunneling of charge carriers with heuristic
expressions for the effective potential $U_{eff}(T,x)=U(x)-E(T,x)$ and
the critical concentration of copper $x_c$.

Part of this work has been financially supported by the Action de
Recherche Concert\'ees (ARC) 94-99/174. S.S. thanks FNRS (Brussels)
for some financial support.

\end{multicols}

\begin{figure}
\epsfxsize=18cm \centerline{\epsffile{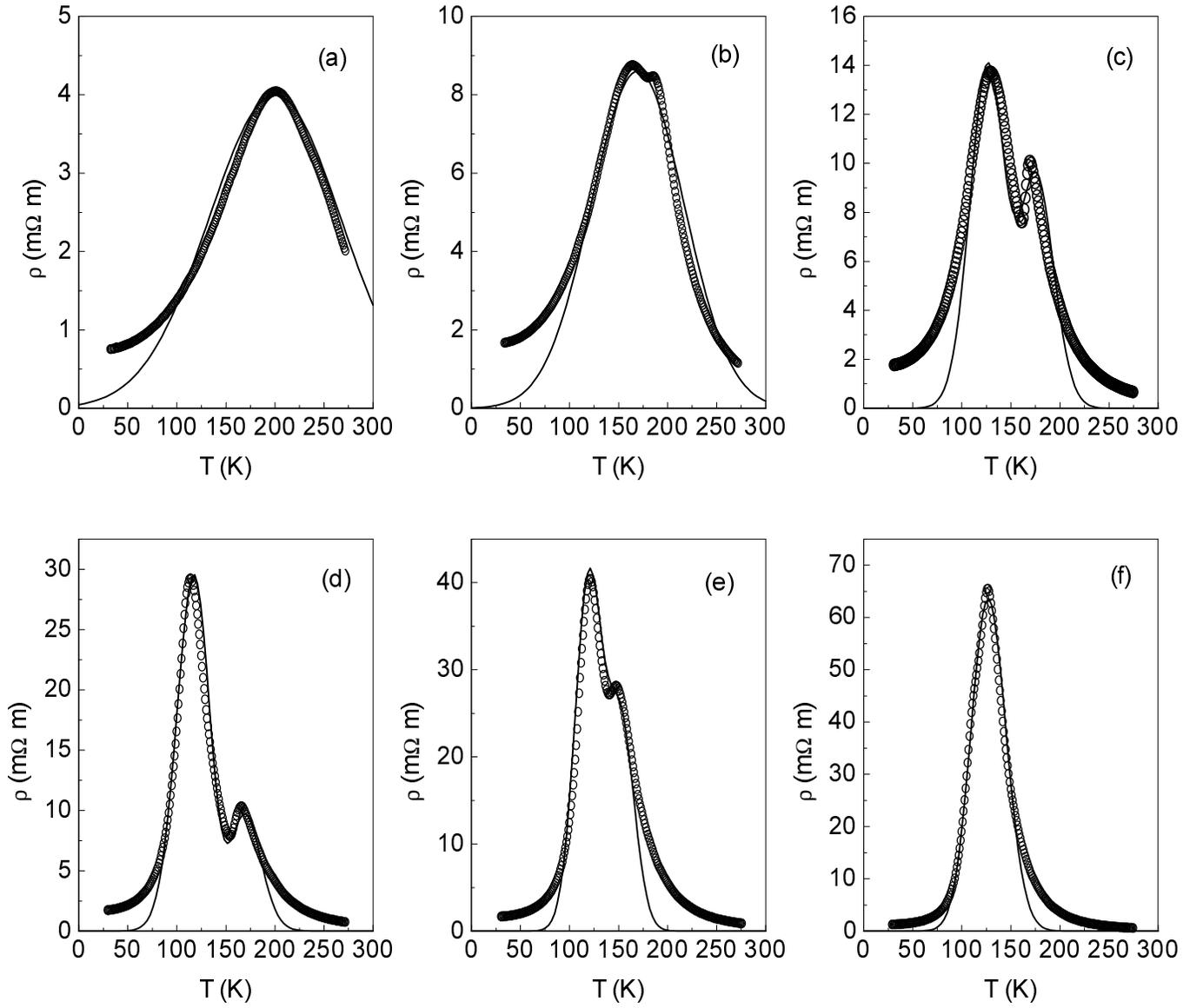}} \caption{Temperature
behavior of the observed resistivity $\rho (T,x)$ in
$La_{0.7}Ca_{0.3}Mn_{1-x}Cu_xO_3$ for different copper content: (a)
$x=0$, (b) $x=0.01$, (c) $x=0.02$, (d) $x=0.03$, (e) $x=0.04$, and (f)
$x=0.05$. The solid lines are the best fits according to
Eqs.(3)-(10).}
\end{figure}

\begin{figure}[htb]
\epsfxsize=17cm \centerline{\epsffile{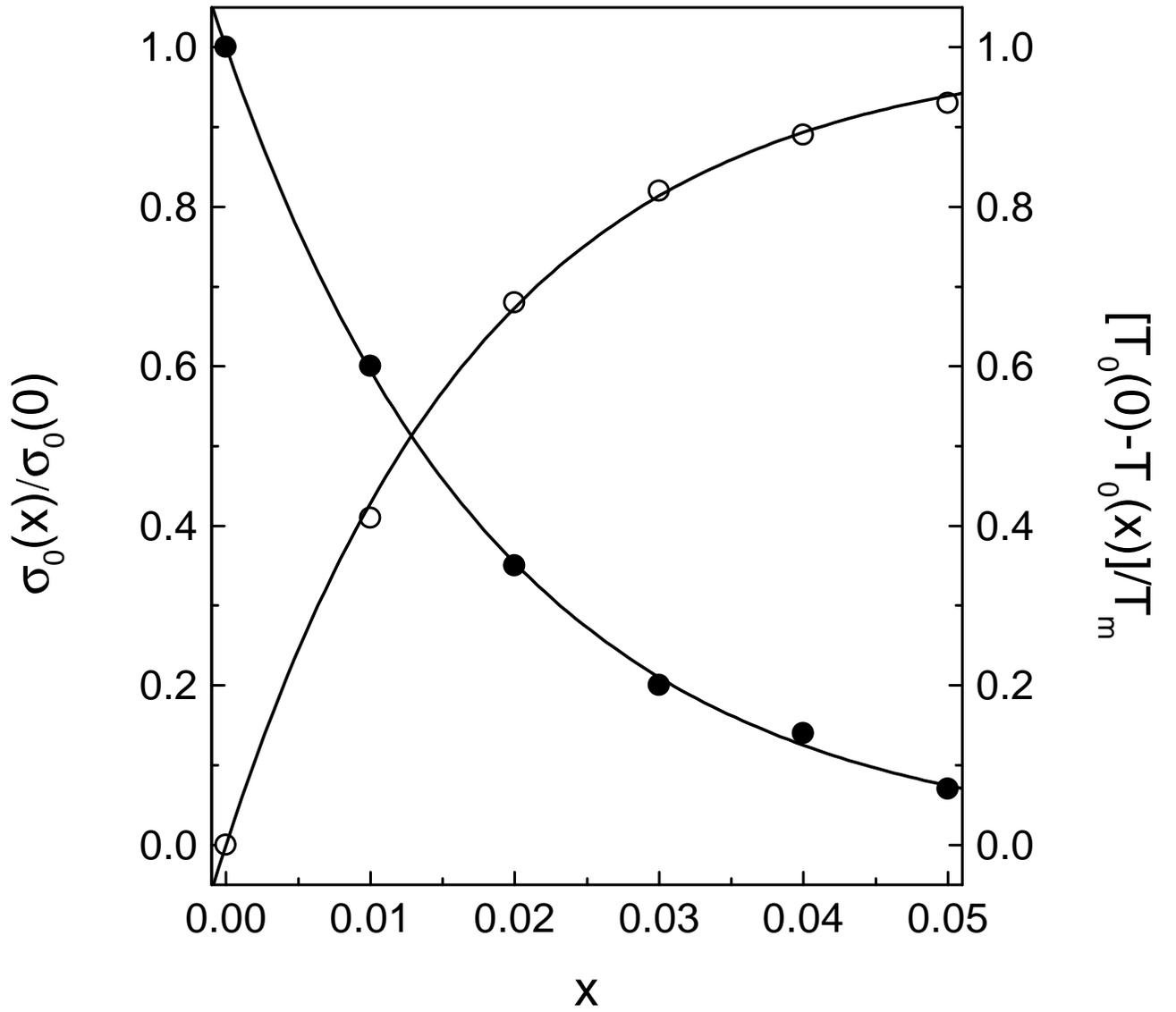}} \caption{The
dependence of the normalized left peak temperature $[T_0(0)-T_0(x)]/
T_m$ (open dots) and conductivity $\sigma _0(x)/\sigma _0(0)$ (solid
dots) on copper doping $x$ in $La_{0.7}Ca_{0.3}Mn_{1-x}Cu_xO_3$. The
solid lines are the best fits according to Eqs.(7) and (8).}
\end{figure}
\end{document}